# Saddle-Node Bifurcation Associated with Parasitic Inductor Resistance in Boost Converters

Chung-Chieh Fang

*Advanced Analog Technology, 2F, No. 17, Industry E. 2nd Rd., Hsinchu 300, Taiwan, Tel: +886-3-5633125 ext 3612*

**Abstract**

Saddle-node bifurcation occurs in a boost converter when parasitic inductor resistance is modeled. Closed-form critical conditions of the bifurcation are derived. If the parasitic inductor resistance is modeled, the saddle-node bifurcation occurs in the voltage mode control or in the current mode control with the voltage loop closed, but not in the current mode control with the voltage loop open. If the parasitic inductor resistance is not modeled, the saddle-node bifurcation does not occur, and one may be misled by the wrong dynamics and the wrong steady-state solutions. The saddle-node bifurcation still exists even in a boost converter with a popular type-III compensator. When the saddle-node bifurcation occurs, multiple steady-state solutions may coexist. The converter may operate with a voltage jump from one solution to another. Care should be taken in the compensator design to ensure that only the desired solution is stabilized. In industry practice, the solution with a higher duty cycle (and thus the saddle-node bifurcation) may be prevented by placing a limitation on the maximum duty cycle.

*Keywords:* DC-DC power conversion, voltage mode control, current mode control, parasitic resistance, saddle-node bifurcation

## 1. Introduction

A DC-DC switching converter is a nonlinear system with rich dynamics. Parasitic *capacitor* resistance, also known as equivalent series resistance (ESR), in a DC-DC converter adds a zero in the dynamics, and it is generally considered in most converter models [1]. Parasitic *inductor* resistance is known to change the steady-state solutions [1]. However, it is generally not included in most models for analysis. A saddle-node bifurcation (SNB) associated with the parasitic inductor resistance in a boost converter under current mode control (CMC) has been reported in [2] by simulation without deriving the bifurcation critical condition. Analysis of SNB may explain some sudden disappearances or jumps of steady-state solutions observed in switching converters [3, 4, 5, 6, 7, 8].

In a boost converter when the parasitic inductor resistance is not modeled, the output voltage ($v_o(D) = v_s/(1-D)$) increases monotonously as $D$ increases, where $v_s$ is the source voltage and $D$ is the duty cycle. One output voltage. has only one corresponding duty cycle (see the dashed line in Fig. 1). However, in a boost converter when the parasitic inductor resistance is modeled, $v_o(D)$ is $\Lambda$-shaped [1, p. 98]. One output voltage has *two* corresponding duty cycles (see the solid line in Fig. 1). The SNB is generally associated with coexistence of multiple solutions [3]. However, such coexistence of multiple solutions in the boost converter has been seldom linked to the SNB. In this paper, the critical conditions are derived for the SNB associated with the parasitic inductor resistance in a boost converter under either CMC or voltage mode control (VMC). The effects of the parasitic inductor resistance on the dynamics are analyzed.

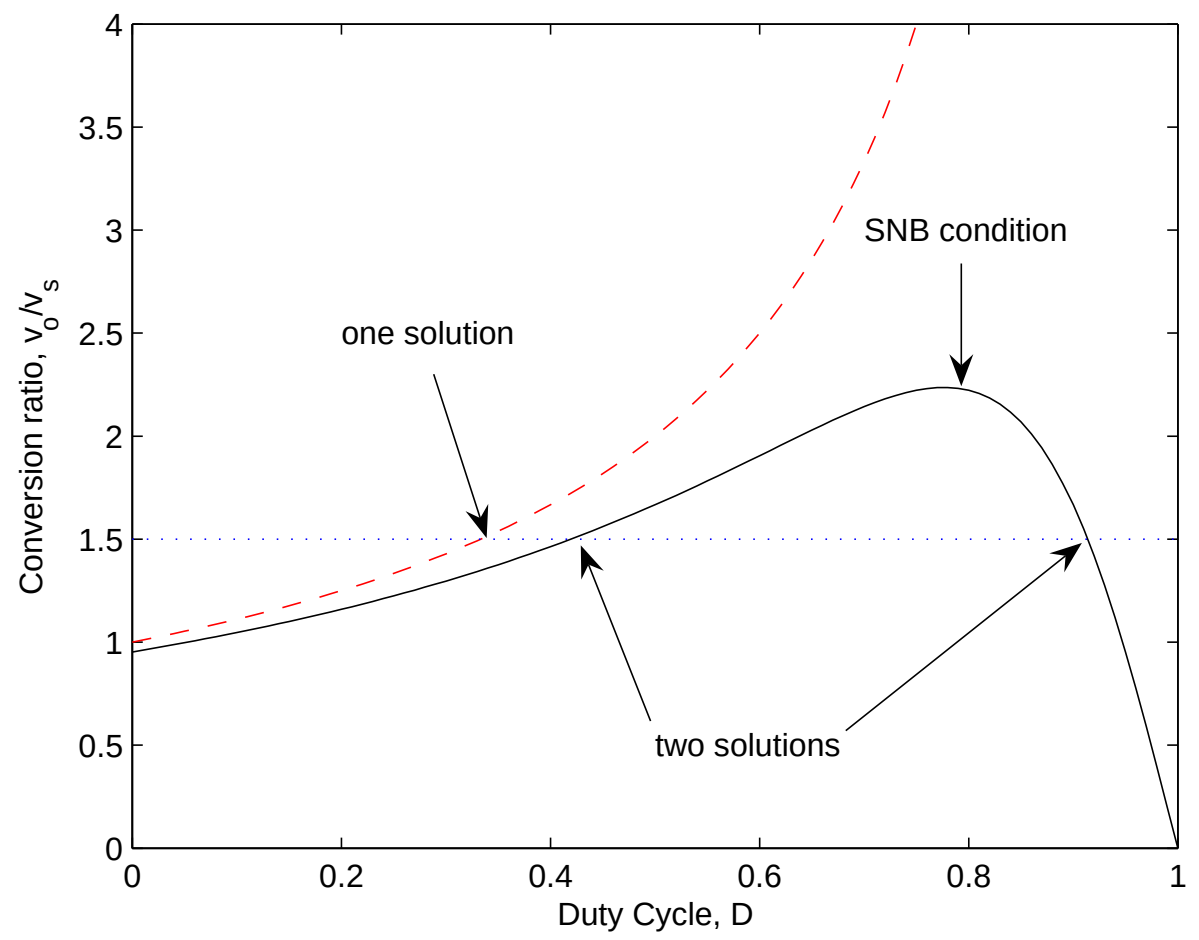

Figure 1: Conversion ratio for a boost converter with zero (dashed line) or nonzero (solid line) parasitic inductor resistance. SNB occurs when two solutions coalesce.

The remainder of the paper is organized as follows. In Section 2, a unified voltage/current mode control model is discussed. In Sections 3 and 4, the saddle-node bifurcations associated with the parasitic inductor resistances in VMC and CMC are analyzed. Conclusions are collected in Section 5.

## 2. Unified VMC/CMC Model

The operation of a DC-DC switching converter under VMC *or* CMC can be described exactly by a unified block diagram model [9] shown in Fig. 2. In VMC or in CMC with a closed voltage loop, the reference signal $v_r$ controls the output voltage $v_o$. In CMC, $v_r$ is denoted as $i_c$, and it controls the peak inductor current $i_L$. In the model, $A_1, A_2 \in \mathbf{R}^{N \times N}$, $B_1, B_2 \in \mathbf{R}^{N \times 2}$, $C, E_1, E_2 \in \mathbf{R}^{1 \times N}$, and $D \in \mathbf{R}^{1 \times 2}$ are constant matrices, where $N$ is the system dimension. Since the output voltage may be discontinuous, let $E = (E_1 + E_2)/2$. Confusion of notations for capacitance $C$ and duty cycle $D$ with the matrices $C$ and $D$ can be avoided from the context. Within a clock period $T$, the dynamics is switched between two stages, $S_1$ and $S_2$ (for continuous conduction mode). Switching occurs when the ramp signal $h(t)$ intersects with the compensator output $y := Cx + Du \in \mathbf{R}$. Denote the ramp amplitude as $V_h$, and denote the switching frequency as $f_s = 1/T$.

## 3. Voltage Mode Control (VMC)

Let the parasitic resistance associated with the inductor be $r$. The parasitic resistance associated with the switch or the diode can be also included in $r$ [1, p. 437]. Without loss of generality, assume that the ESR is small enough such that the zero associated with the ESR is at infinity. As discussed below, SNB is associated with the steady-state solution, and the ESR has a minor effect on the steady state and on SNB.

Let the state be $x = (i_L, v_C)'$, where $i_L$ is the inductor current, and $v_C$ is the capacitor voltage. For small ESR, $v_C \approx v_o$. Consider VMC with a proportional feedback gain $k_p$ as shown in Fig. 3.

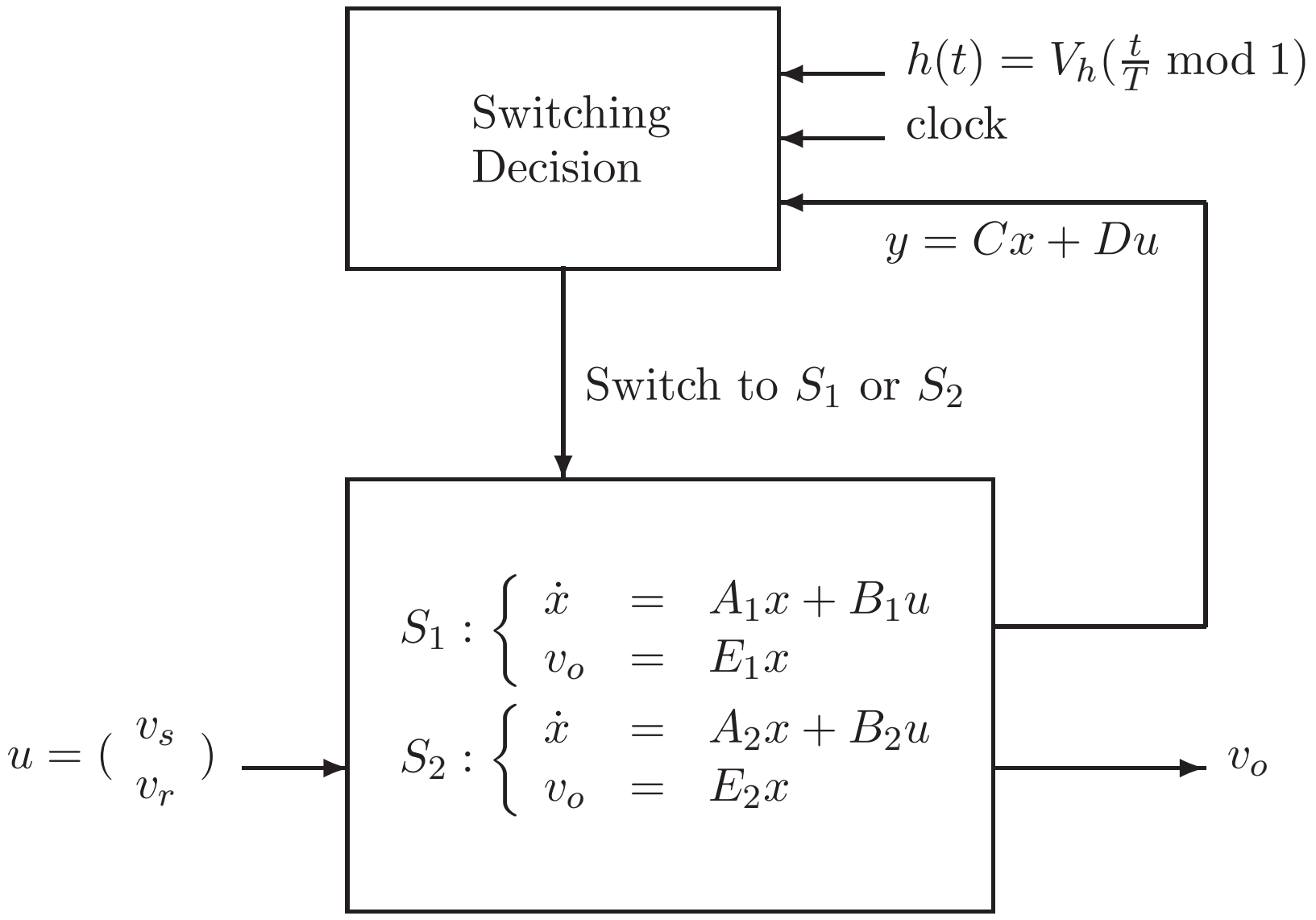

Figure 2: Block diagram model for switching converter.

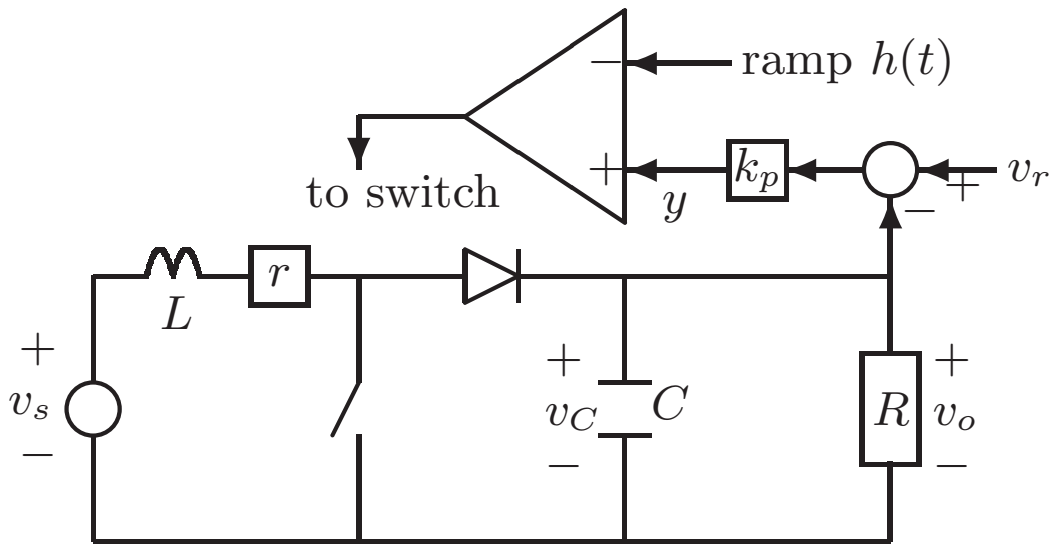

Figure 3: A boost converter under PVMC.

This control scheme is called PVMC here. One has $y = k_p(v_r - v_C)$ and

$$
\begin{aligned}
A_1 &= \begin{bmatrix} \frac{-r}{L} & 0 \\ 0 & \frac{-1}{RC} \end{bmatrix}, \quad A_2 = \begin{bmatrix} \frac{-r}{L} & \frac{-1}{L} \\ \frac{1}{C} & \frac{-1}{RC} \end{bmatrix} \\
B_1 &= B_2 = \begin{bmatrix} \frac{1}{L} & 0 \\ 0 & 0 \end{bmatrix} \\
C &= \begin{bmatrix} 0 & -k_p \end{bmatrix}, \quad D = \begin{bmatrix} 0 & k_p \end{bmatrix} \\
E_1 &= E_2 = E = \begin{bmatrix} 0 & 1 \end{bmatrix}
\end{aligned}
\tag{1}
$$

### *3.1. Coexistence of Multiple Steady-State Solutions*

Let $\eta = r/R$, $A = DA_1 + (1-D)A_2$, and $B = DB_1 + (1-D)B_2$. In the state-space average (SSA) model [10], the power stage dynamics is $\dot{x} = Ax + Bu$, and the steady-state solution is $-A^{-1}Bu := X$. From (1), one has

$$
X := \begin{bmatrix} I_L(D) \\ V_C(D) \end{bmatrix} = \frac{v_s}{\eta + (1-D)^2} \begin{bmatrix} \frac{1}{R} \\ 1 - D \end{bmatrix}
\tag{2}
$$

where $V_C(D)$ is $\Lambda$-shaped and has a maximum of $v_s/2\sqrt{\eta}$ at

$$D = 1 - \sqrt{\eta} := D_S \tag{3}$$

Note that $I_L(D)$ increases monotonously as $D$ increases. If $r = 0$, $V_C(D)$ also increases monotonously as $D$ increases (Fig. 1). However, if $r > 0$, $v_o(D)$ is $\Lambda$-shaped (Fig. 1). As discussed below, the shapes of $V_C(D)$ and $I_L(D)$ determine whether multiple solutions coexist and thus whether the SNB occurs.

Let $\kappa = k_p/V_h$. In steady state, $DV_h = y = k_p(v_r - V_C(D))$, rearranged as

$$v_r = \frac{D}{\kappa} + V_C(D) := v_r(D) \tag{4}$$

**[Case 1: Large** $k_p$**]** For a large $k_p$, $D/\kappa$ can be ignored. Since $V_C$ has a maximum of $v_s/2\sqrt{\eta}$, two steady-state solutions exist if $v_r < v_s/2\sqrt{\eta} := v_r^*$. One of the solutions has $D > D_S$, and the other has $D < D_S$. SNB occurs when $v_r = v_r^*$, or equivalently when $D = D_S$. No solution exists if $v_r > v_r^*$.

**[Case 2: Small** $k_p$**]** For a small $k_p$, a more accurate calculation of $v_r^*$ from (4) is

$$v_r^* = v_r(D_S) = \frac{v_s}{2\sqrt{\eta}} + \frac{1-\sqrt{\eta}}{\kappa} \tag{5}$$

Compared with $v_r^* = v_s/2\sqrt{\eta}$ for a large $k_p$, The critical $v_r^*$ for a small $k_p$ has an additional term $(1 - \sqrt{\eta})/\kappa$.

*3.2. Linearized Dynamics*

Using a hat $\hat{}$ to denote small perturbations, (e.g., $\hat{x} = x - X$), the linearized SSA dynamics [10] is

$$\dot{\hat{x}} = A\hat{x} + (A_1 - A_2)X\hat{D} \tag{6}$$

The control-to-output transfer function is

$$G_{vd}(s) = \frac{\hat{v}_o(s)}{\hat{D}(s)} = E(sI - A)^{-1}(A_1 - A_2)X \tag{7}$$

The closed-loop dynamics is $\kappa G_{vd}(s)/(1+\kappa G_{vd}(s))$. Arrange the closed-loop characteristic equation as $s^2 + c_1 s + c_0 = 0$, where the coefficients $c_1$ and $c_0$ are [1, p. 441]

$$c_1 = \frac{r}{L} + \frac{1}{RC} - \frac{\kappa I_L(D)}{C} \tag{8}$$

$$c_0 = \frac{1}{LC}(\eta + (1-D)^2 + \kappa R I_L(D)((1-D)^2 - \eta)) \tag{9}$$

Both $c_1$ and $c_0$ are functions of $D$. One can prove that $c_1 > 0$ if

$$D < D_H := 1 - \sqrt{\frac{\kappa v_s}{\frac{rRC}{L} + 1} - \eta} \tag{10}$$

and $c_0 > 0$ if

$$D < D_S := 1 - \sqrt{\sqrt{(2\eta + \frac{\kappa v_s}{4})\kappa v_s} - \eta - \frac{\kappa v_s}{2}} \tag{11}$$

$$\approx 1 - \sqrt{\eta} \text{ (for a large } \kappa) \tag{12}$$

It will be shown later that $D_H$ and $D_S$ are the critical duty cycles for the Hopf bifurcation and the SNB, respectively. Both $D_H$ and $D_S$ are functions of $\eta = r/R$, and $D_S$ can be approximated as a function of only $\eta$. This paper focuses on SNB. Other bifurcations are discussed only when necessary.

*3.3. Bifurcation Critical Conditions*

The poles are the roots of the characteristic equation. From the Routh-Hurwitz stability criterion, there are two unstable poles if $c_1 < 0$ and $c_0 > 0$, one unstable pole if $c_0 < 0$, or two stable poles if $c_1 > 0$ and $c_0 > 0$. The pole location is a function of $D$.

*3.3.1. Hopf bifurcation*

When the Hopf bifurcation occurs, two poles cross the imaginary axis. Then, $c_1 = 0$ and $c_0 > 0$. The bifurcation critical conditions are $D = D_H$ and $v_r = v_r(D_H)$.

From (10), $D_H < 0$ if

$$\kappa > \frac{1+\eta}{v_s}(\frac{rRC}{L} + 1) \tag{13}$$

and the Hopf bifurcation does not occur because the duty cycle is never less than zero. For $r = 0$, the condition (13) becomes $\kappa > 1/v_s$.

*3.3.2. Saddle-node bifurcation*

When SNB occurs, a pole is zero and $c_0 = 0$. The bifurcation critical conditions are $D = D_S$ and $v_r = v_r^*$. Note that the condition (3) (for coexistence of multiple solutions) and the condition (12) (for occurrence of SNB) are the same, indicating that coexistence of multiple solutions is associated with SNB.

As $D$ increases, the Hopf bifurcation occurs earlier than the SNB does if $D_H < D_S$. From (10) and (12), $D_H < D_S$ if

$$\kappa > \frac{2\eta}{v_s}(\frac{rRC}{L} + 1) \tag{14}$$

which is generally true for a converter with a large $k_p$.

*3.4. No Saddle-Node Bifurcation if $r = 0$*

For $r = 0$, one can prove that SNB does not occur in two ways. First, from (9), $c_0 > 0$ if $r = 0$. The pole is never zero, and SNB does not occur. Second, the right side of (4) increases monotonously as $D$ increases if $r = 0$. Given a value of $v_r$, only one solution exists. Coexistence of multiple solutions as in SNB does not occur.

For $r = 0$ (thus $c_0 > 0$), the pole stabilities depend on $c_1$. If $c_1 > 0$, the two poles are in the left half plane (LHP) and stable. If $c_1 < 0$, the two poles are in the right half plane (RHP) and unstable. If $c_1 = 0$, the Hopf bifurcation occurs. In all cases, it never occurs that one pole is stable and the other is unstable.

**Example 1.** (*PVMC.*) Consider a boost converter under PVMC with $k_p = 2$. The converter parameters are $v_s = 3$ V, $V_h = 1$, $f_s = 600$ kHz, $L = 1$ $\mu$H, $C = 100$ $\mu$F, $R = 2$ $\Omega$, and parasitic inductor resistance $r = 0.1$ $\Omega$. One has $\eta = r/R = 0.05$ and $\kappa = k_p/V_h = 2$. For comparison, the bifurcation diagrams for $r = 0.1$ and $r = 0$ are shown in Fig. 4. For $r = 0.1$, the Hopf bifurcation occurs at $v_r = 4.92$ and SNB occurs at $v_r = 7.1$. For $r = 0$, no bifurcation occurs.

Note that the focus of this paper is on the $T$-periodic or DC solutions ($D = 1$ or 0). For $r = 0.1$, let $v_r = 7$, for example. The two unstable $T$-periodic solutions are shown in Fig. 5. One solution has $D = 0.74$, and the other has $D = 0.81$, agreed with Fig. 4(a). Other stable attractors may coexist with the two unstable solutions.

In this particular example, besides $T$-periodic solutions, there actually exists a DC solution ($D = 1$ not shown in Fig. 4, with one switch being always on because $y(t) = k_p(v_r - v_C) = k_p v_r > h(t)$) with $(i_L, v_C) = (v_s/r, 0) = (30, 0)$. Such a DC solution does not exist if $r = 0$ because it requires $i_L = v_s/r = \infty$. Therefore, the converter is actually *bistable* (with the DC solution *and* the $T$-periodic solution) for $4 < v_r < 4.92$ and monostable (with only the DC solution) for $v_r > 4.92$. From Fig. 4(b), if $v_r$ increases a little from 4.92, $v_c$ may "jump" from 4.7 (with the $T$-periodic solution) to 0 (with the DC solution as the only solution). Such a jump is typical in a system with SNB [3].

With different converter parameters, the attractor may be quasi-periodic [3], associated with the Hopf bifurcation. Other attractors associated with the border-collision bifurcation [11] may also exist when $y(t)$ of the attractor is out of the bounds of $h(t)$. These non-$T$-periodic attractors are not shown in the bifurcation diagrams to prevent detraction of the focus on SNB.

For $r = 0.1$, the coefficients $c_0$ and $c_1$ as functions of $D$ are shown in Fig. 6.The pole loci are shown in Fig. 7. It shows that SNB occurs when two unstable solutions coalesce because each solution has a RHP pole. From (5) and (11), $v_r^* = 7.1$ and $D_S = 0.78$, the prediction of the SNB point agrees closely with the bifurcation diagram (Fig. 4(a)). From (10) and (4), $D_H = 0.51$ and $v_r(D_H) = 5.36$, and a small error exists in predicting the Hopf bifurcation point. The error is due to averaging. If the switching frequency increases to 6 MHz, instead of 600 kHz, the Hopf bifurcation occurs at $v_r = 5.32$, agreed closely with the prediction. Since this paper focuses on SNB, the discussion on the averaging error for the Hopf bifurcation is omitted.

For $r = 0.1$, SNB occurs at $v_r = v_r^* = 7.1$ and $D_S = 0.78$. For $v_r > v_r^*$, no solution exists. Two solution branches are created in the bifurcation diagram for $v_r < v_r^*$. The first branch has $D > D_S = 0.78$. Since this branch has $c_0 < 0$, and $c_1 < 0$, one pole is in LHP and the other pole is in RHP. The first branch is unstable.

The second branch has $D < D_S = 0.78$ and $c_0 > 0$. Since $0 < D_H < D_S$, the lower solution branch has two sections, separated by the Hopf bifurcation point at $v_r = 4.92$, as shown in Fig. 4(a). For $v_r < 4.92$ ($c_1 > 0$ and $c_0 > 0$), the two poles are in LHP, and this section is stable. At $v_r = 4.92$ ($c_1 = 0$ and $c_0 > 0$), the Hopf bifurcation occurs. For $v_r > 4.92$ ($c_1 < 0$ and $c_0 > 0$), the two poles are in RHP, and this section is unstable.

The value of $D_H$ (and thus the length of stable section of the lower solution branch in the bifurcation diagram shown in Fig. 4(a)) may be adjusted by $\kappa$. From (10), $D_H$ decreases if $\kappa$ increases. If $\kappa$ is large enough such that $D_H < 0$, the lower solution branch is unstable. Similarly, $D_H$ increases if $\kappa$ decreases. If $\kappa$ is small enough such that $D_H > D_S$, the lower solution branch is stable, connecting with the upper solution branch at the SNB point.

By comparing Figs. 4(a)(b) and (c)(d), one can see significant differences *whether* $r$ is modeled. When $r$ is modeled, two solutions (branches) exist, and one solution is stable for $v_r < 4.92$. When $r$ is not modeled, only one solution exists and it is unstable. From (10), $D_H$ increases if $\kappa$ decreases. For $r = 0$, from (13), $\kappa$ needs to decrease below $1/v_s = 0.5$ to have a stable solution.

Therefore, if $r$ is not modeled, one may be misled by the bifurcation diagrams shown in Figs. 4(c)(d) that no stable solution exists and no bifurcation occurs. A small parasitic inductor resistance, always existing in reality, may have great effects. As shown in Figs. 4(a)(b), a stable solution does exist for $v_r < 4.92$. Furthermore, SNB occurs and an unstable solution actually coexists with the stable solution. Care should be taken in the compensator design because different compensator parameters may result in a *different solution* being stabilized. What the compensator stabilizes may be an undesired solution [6]. □

### *3.5. No Saddle-Node Bifurcation in Critical Mode*

Let $K = 2L/RT$. To operate in the critical mode on the continuous/discontinuous conduction mode (CCM/DCM) boundary, it is required [10] that

$$K = D(1 - D)^2 := K_{\text{crit}} \tag{15}$$

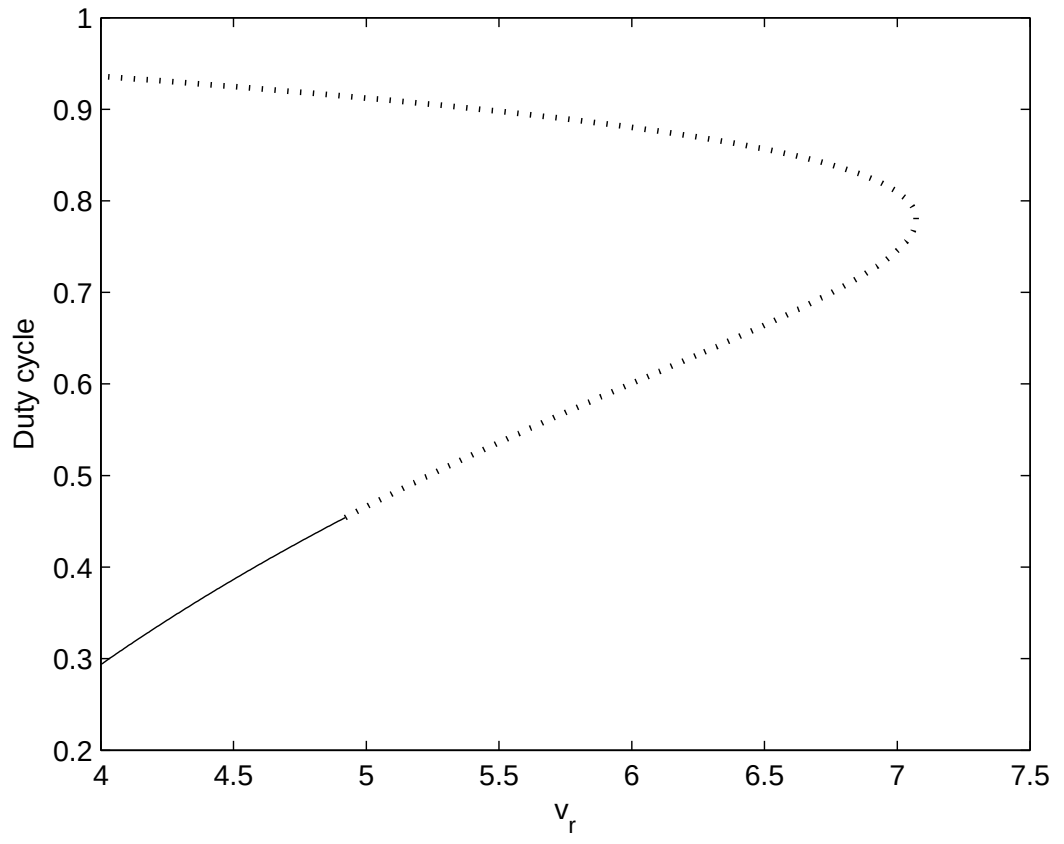

(a) $r = 0.1$, SNB occurs at $v_r = 7.1$ (Hopf bifurcation occurs at $v_r = 4.92$).

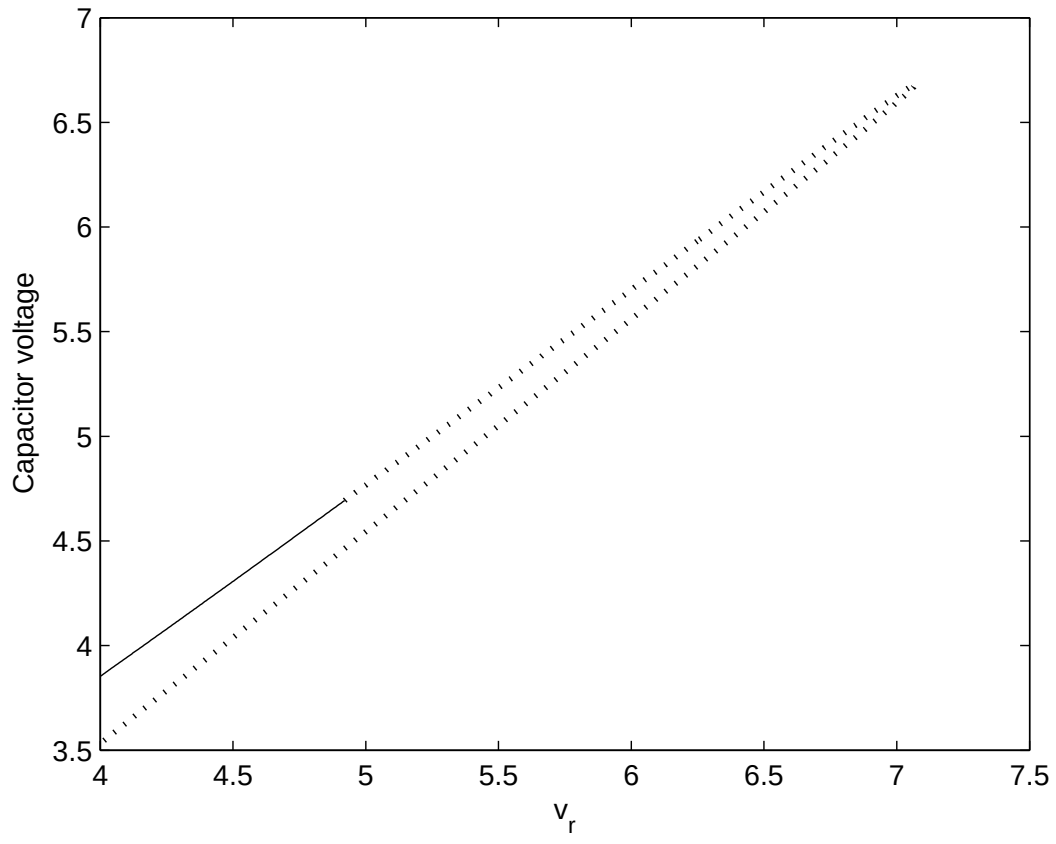

(b) $r = 0.1$, SNB occurs at $v_r = 7.1$.

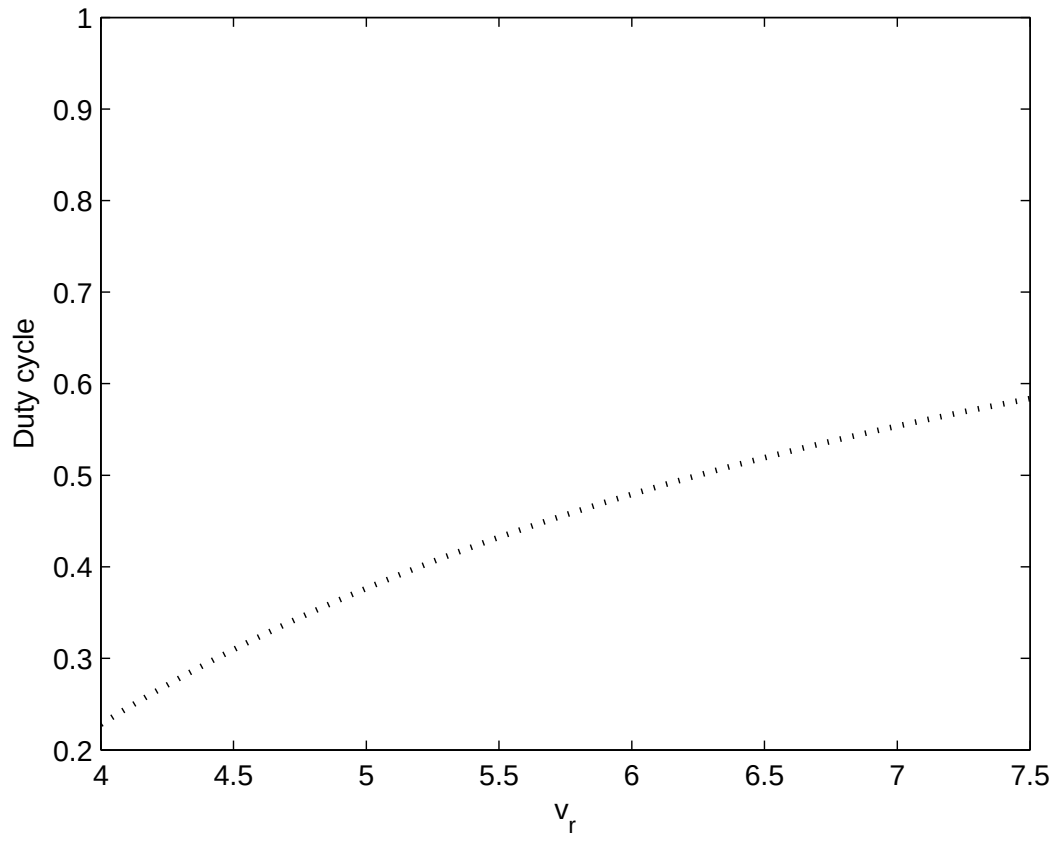

(c) $r = 0$, no SNB occurs.

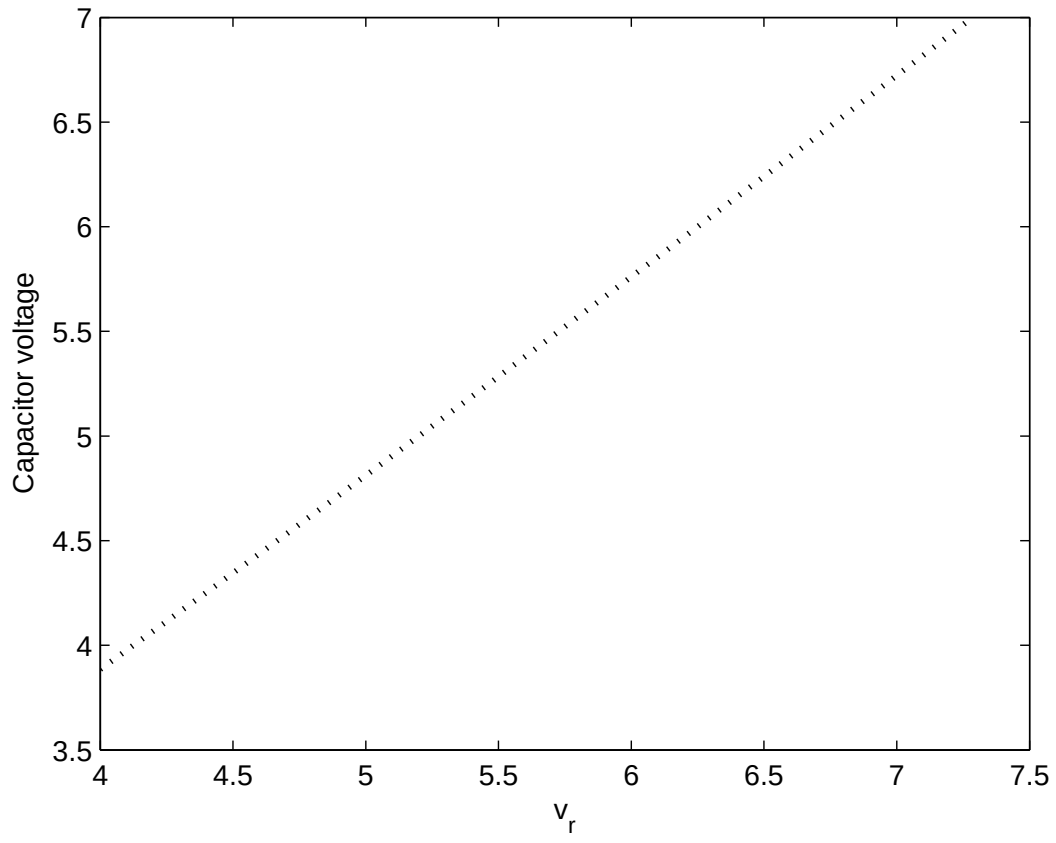

(d) $r = 0$, no SNB occurs.

Figure 4: Bifurcation diagrams showing stable (solid) and unstable (dotted) solutions. .

To have SNB, $D = 1 - \sqrt{\eta} = 1 - \sqrt{KrT/2L}$ which is equivalent to

$$K = \frac{2L}{rT}(1 - D)^2 := K^* \tag{16}$$

Generally, $2L/rT > D$ and $K^* > K_{\text{crit}}$, and SNB does not occur in the VMC boost converter in the critical mode.

### *3.6. VMC with a type-III compensator*

As discussed above, SNB is associated with the steady-state solutions. SNB occurs when the two solutions coalesce. With the *same* parasitic inductor resistance $r$, the (multiple) steady-state solutions are the *same* as in PMVC shown in (2). The Hopf bifurcation may be prevented by the type-III compensator, but SNB *always* occurs if $r > 0$.

Let the type-III compensator be

$$G_c(s) = \frac{K_c(1 + \frac{s}{z_1})(1 + \frac{s}{z_2})}{s(1 + \frac{s}{p_1})(1 + \frac{s}{p_2})} \tag{17}$$

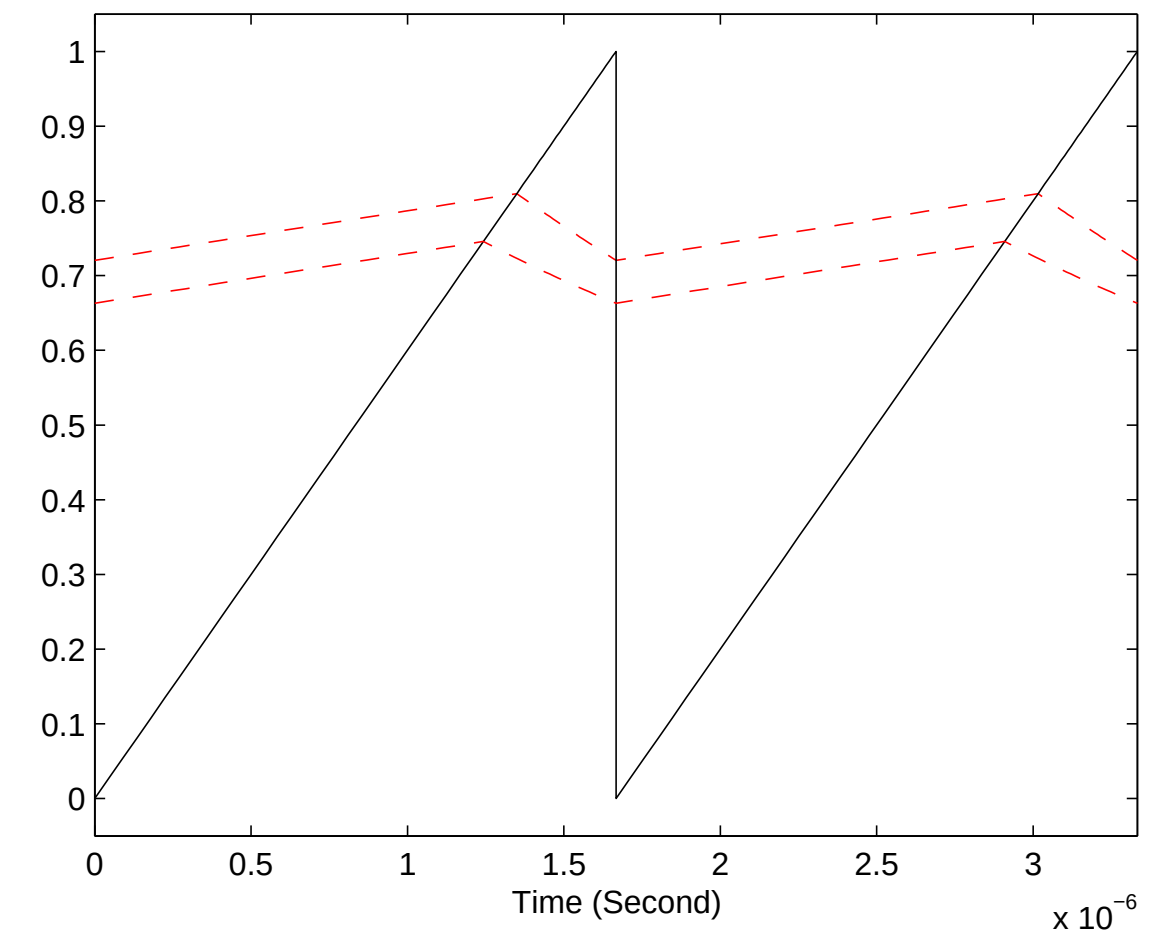

Figure 5: Ramp $h(t)$ (solid line) and two coexisting unstable $T$-periodic solutions $y^0(t)$ (dashed line), $v_r = 7$.

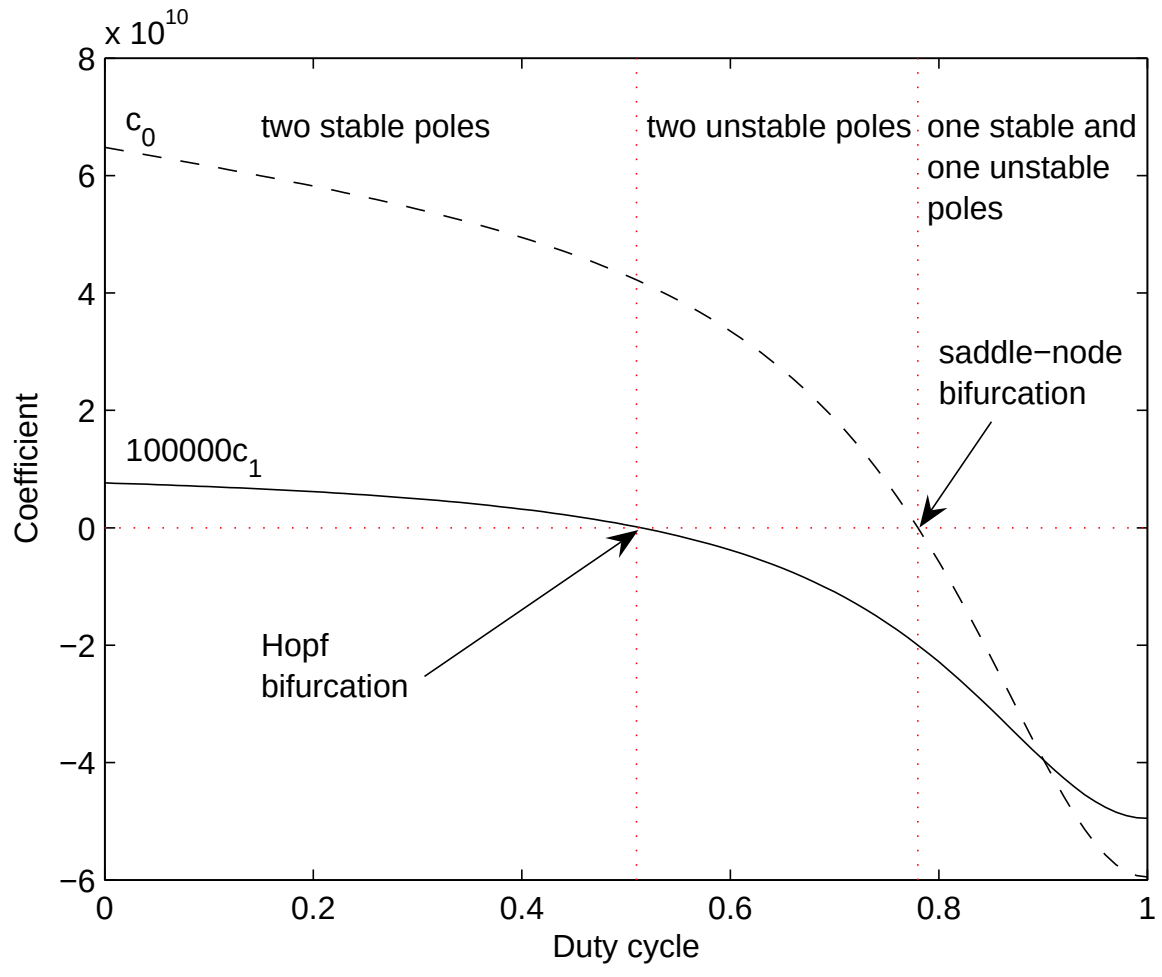

Figure 6: Coefficients $c_0$ and $c_1$ as functions of $D$.

The compensator output is

$$y = v_r + G_c(s)(v_r - v_o) \tag{18}$$

With a high feedback gain, the SNB critical condition is $D = D_S$ and $v_r \approx v_C(D)$ as shown in (2) because the the steady-state solutions are the same.

**Example 2.** (*VMC with a type-III compensator.*) Consider a VMC boost converter [12] with a type-III compensator. The same parameters as in [13, p. 446] are used: $f_s = 1/T = 300$ kHz, $L = 46.6$ $\mu$H, $C = 3$ mF, $R = 23$ $\Omega$, ESR $R_c = 18$ m$\Omega$, $v_s = 10$ V, $V_h = 2$ V, $K_c = 35.59$, $z_1 = 556$, $z_2 = 549$, $p_1 = 25510$, and $p_2 = 19495$. Let $r = 0.6$ $\Omega$, then $\eta = r/R = 0.0261$. One expects that SNB occurs at $D_S = 1 - \sqrt{\eta} = 0.84$. A smaller value of $r$ (or $\eta$) leads to a larger critical duty cycle. As long as $r > 0$, SNB always exists.

Let the bifurcation parameter be $v_r$. The bifurcation diagram is shown in Fig. 8, with SNB

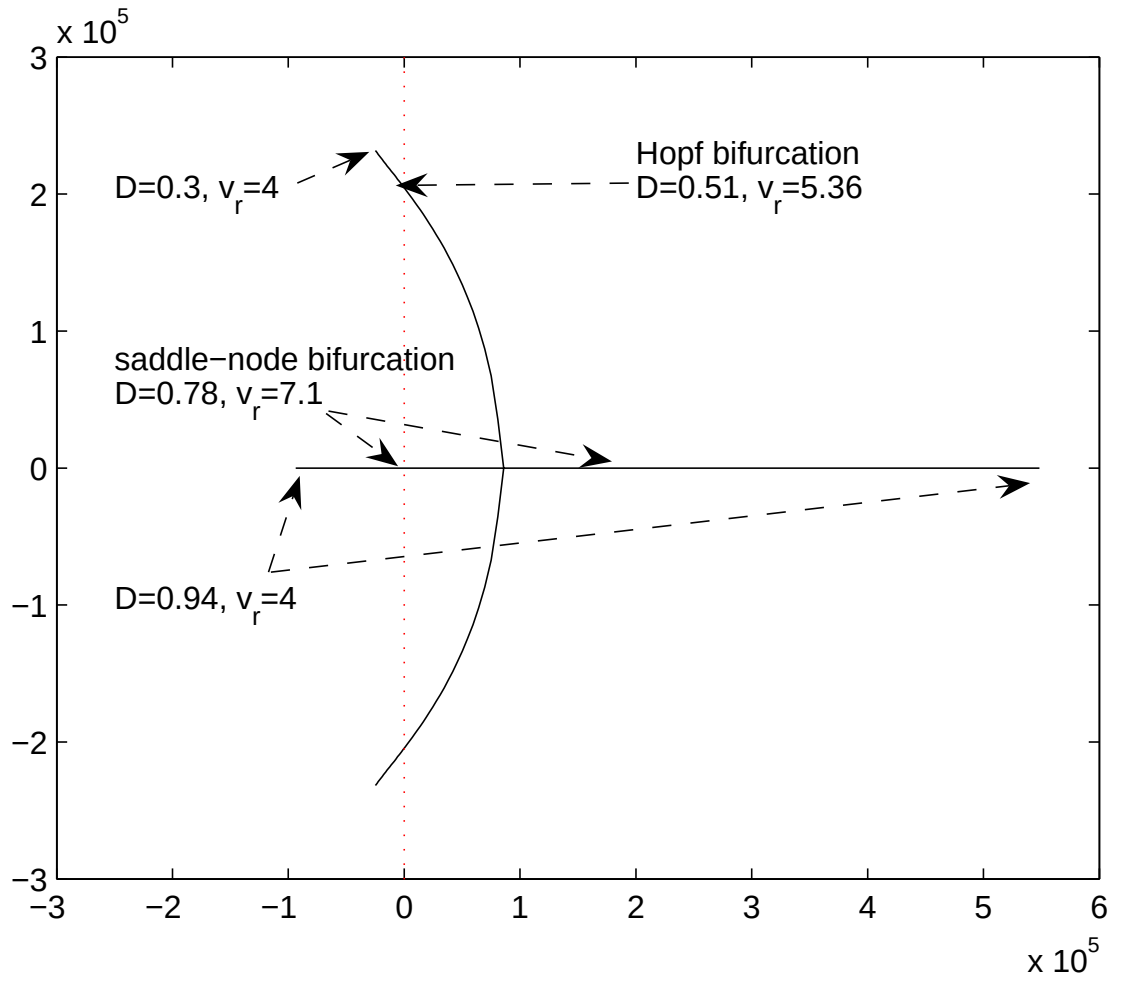

Figure 7: Pole loci as $D$ or $v_r$ varies.

occurring exactly at $D_H = 0.84$ and $v_r = v_C(D_H) = 31$ as expected. The bifurcation diagram is very similar to Fig. 4(a), where a *simple* PVMC is used. In Fig. 4(a), $\eta = r/R = 0.05$, whereas in Fig. 8, $\eta = 0.0261$. For the same $\eta$, using either PVMC or a type-III compensator, the *steady state* of the output voltage in the bifurcation diagram would be the same. The only difference is the stability of the steady-state solution. In Fig. 4(a), SNB occurs when two unstable solutions coalesce, whereas in Fig. 8, SNB occurs when an unstable solution and a stable one coalesce, because a type-III compensator improves the the stability of one solution, but does not affect the steady state of the multiple solutions.

Take $v_r = 30.3$, for example. Two $T$-periodic solutions coexist as shown in Figs. 9 and 10. One is stable with $D = 0.8$ and the other is unstable with $D = 0.87$. The phase portraits of the both solutions in a *single* plot are omitted because the ripple amplitudes are small and the two solutions are apart. Instead, the time-domain plots as in Figs. 9 and 10 are shown.

As in Example 1, there actually exists another DC solution ($D = 1$ because $y = v_r + G_c(0)(v_r - v_0) = (1 + G_c(0))v_r > h(t)$) with $(i_L, v_C) = (v_s/r, 0) = (16.7, 0)$. From Fig. 8, if $v_r$ increases a little from 31, the converter may operate with a "jump" from $D = 0.84$ (with the $T$-periodic solution) to $D = 1$ (with the DC solution as the only solution). Refer to Example 1 for similar discussion.

As discussed above, SNB always exists in the VMC boost converter as long as $r > 0$, even with a type-III compensator. This example also illustrates that even a converter with a simple PVMC controller as in Example 1 deserves careful study because it still provides some insights to design a more complicated type-III controller. □

## 4. Current Mode Control (CMC)

In CMC, two cases are considered depending on whether the voltage loop is closed.

### *4.1. Open Voltage Loop: No Saddle-Node Bifurcation*

First, let the voltage loop open as shown in Fig. 11. A current control signal $i_c$ controls the peak inductor current. From (1), the inductor current ripple $\Delta I_L = (v_s - rI_L)DT/L$. The peak inductor current $I_{peak} = I_L + \Delta I_L/2$ increases monotonously as $D$ increases. Therefore, given a current control signal, only one solution exists and SNB does not occur.

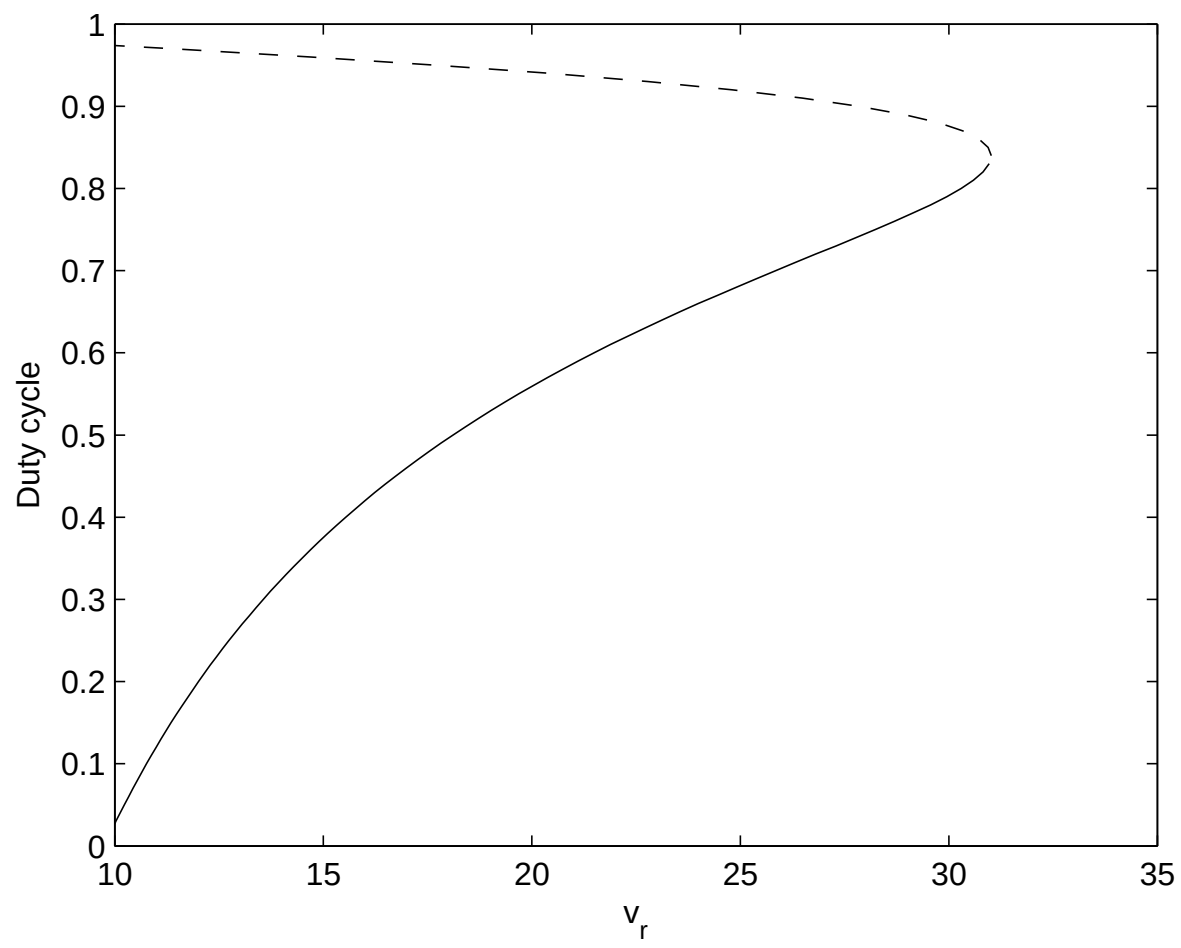

Figure 8: Bifurcation diagram showing occurrence of SNB at $D = 0.84$.

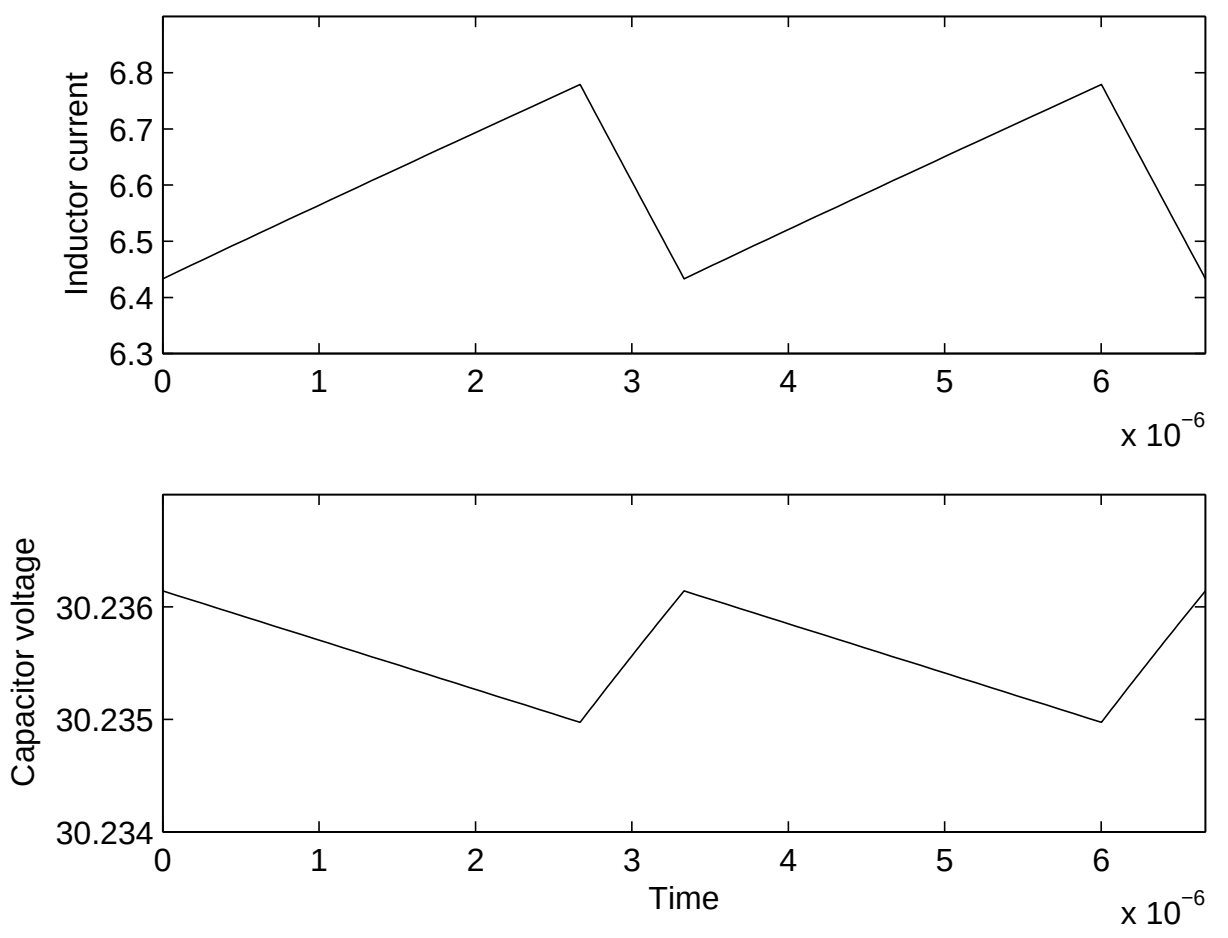

Figure 9: Stable $T$-periodic solution with $D = 0.8$ and $v_r = 30.3$.

*4.2. Closed Voltage Loop: Saddle-Node Bifurcation Occurs if* $r > 0$

Next, close the voltage loop. The voltage loop output controls the peak inductor current. Without loss of generality, let the voltage loop has a proportional feedback gain $k_p$. In terms of Fig. 11, one has $i_c = k_p(v_r - v_c)$. In steady state, $I_{peak} = k_p(v_r - V_C(D))$, which is rearranged as

$$v_r = \frac{I_{peak}}{k_p} + V_C(D) \tag{19}$$

Similar analysis as in VMC (see (4)) can be applied. For a large $k_p$, $I_{peak}/k_p$ can be ignored, and SNB occurs when $v_r^* = v_s/2\sqrt{\eta}$, or equivalently, when $D = D_S = 1 - \sqrt{\eta}$. For a small $k_p$, based on (19), the bifurcation point $D_S$ is larger than $1 - \sqrt{\eta}$ because $I_{peak}$ has a large value at high duty cycle.

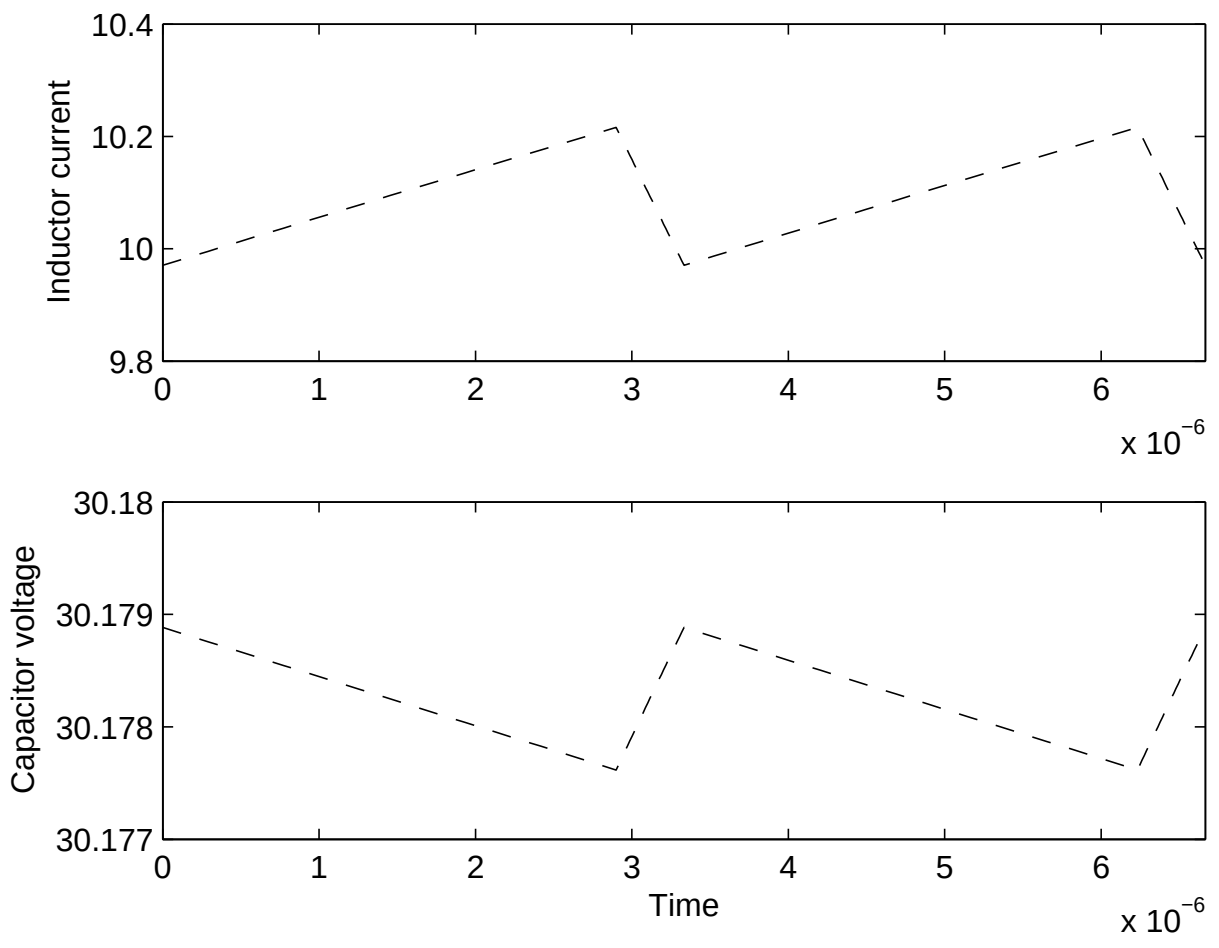

Figure 10: Unstable $T$-periodic solution with $D = 0.87$ and $v_r = 30.3$.

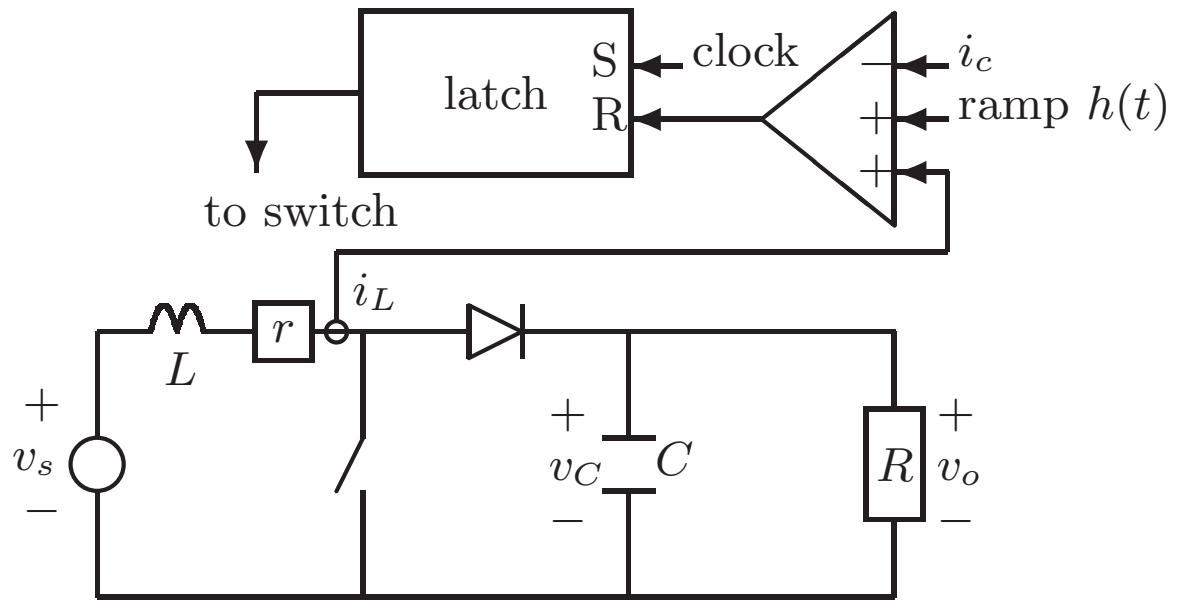

Figure 11: A boost converter under current-mode control.

For $r = 0$, the right side of (19) increases monotonously as $D$ increases. Given a value of $v_r$, only one solution exists, and SNB does not occur. Therefore, SNB does not occur in CMC if $r = 0$, whereas it does occur in CMC with the voltage loop closed if $r > 0$.

**Example 3.** (*CMC with the voltage loop closed.*) Consider a boost converter under CMC with no ramp compensation ($V_h = 0$). The voltage loop has a feedback gain $k_p = 2$. The converter parameters are the same as in Example 1.

For comparison, the bifurcation diagrams for $r = 0.1$ and $r = 0$ are shown in Fig. 12. For $r = 0.1$, SNB occurs at $D_S = 0.91$. Two $T$-periodic solutions exist for $15.5 < v_r < 17.7$. Compared with Example 1, $D_S$ is larger because here $k_p$ is small. For $r = 0$, only one $T$-periodic solution exists. Both the bifurcation diagrams show occurrence of period-doubling bifurcation (PDB) around $D = 0.5$, typical in CMC. For example, Let $r = 0.1$ and $v_r = 8.4$. The the $T$-periodic solution is unstable with period-doubling. The $2T$-periodic $i_L$ and the voltage loop output are shown in Fig. 13. If $r$ is not modeled, one may be misled by Fig. 12(b) that the converter is stable with $v_r = 8.4$. By comparing Figs. 12(a) and 12(b), one also sees significant differences whether $r$ is modeled. SNB occurs in Fig. 12(a), whereas it does not occur in Fig. 12(b). □

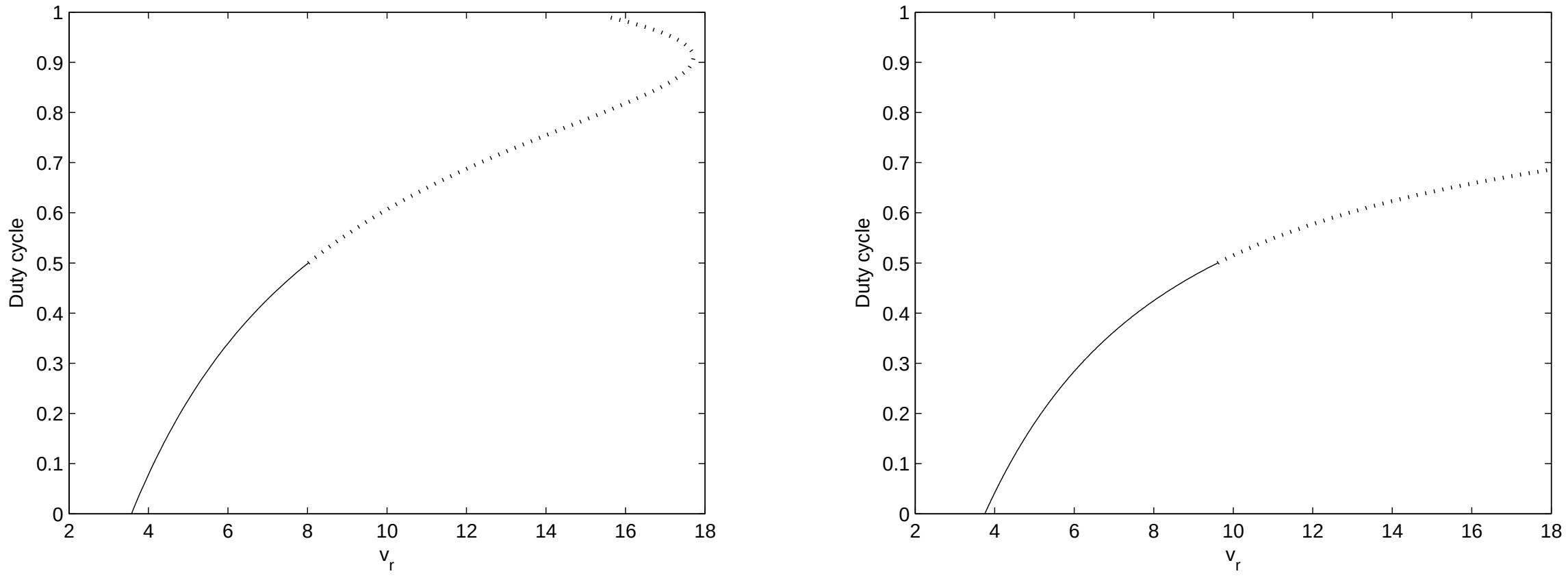

(a) $r = 0.1$, SNB occurs at $v_r = 17.71$ (PDB occurs at $v_r = 8.2$).

(b) $r = 0$, no SNB occurs (PDB occurs at $v_r = 9.4$).

Figure 12: Bifurcation diagrams showing stable (solid) and unstable (dotted) solutions.

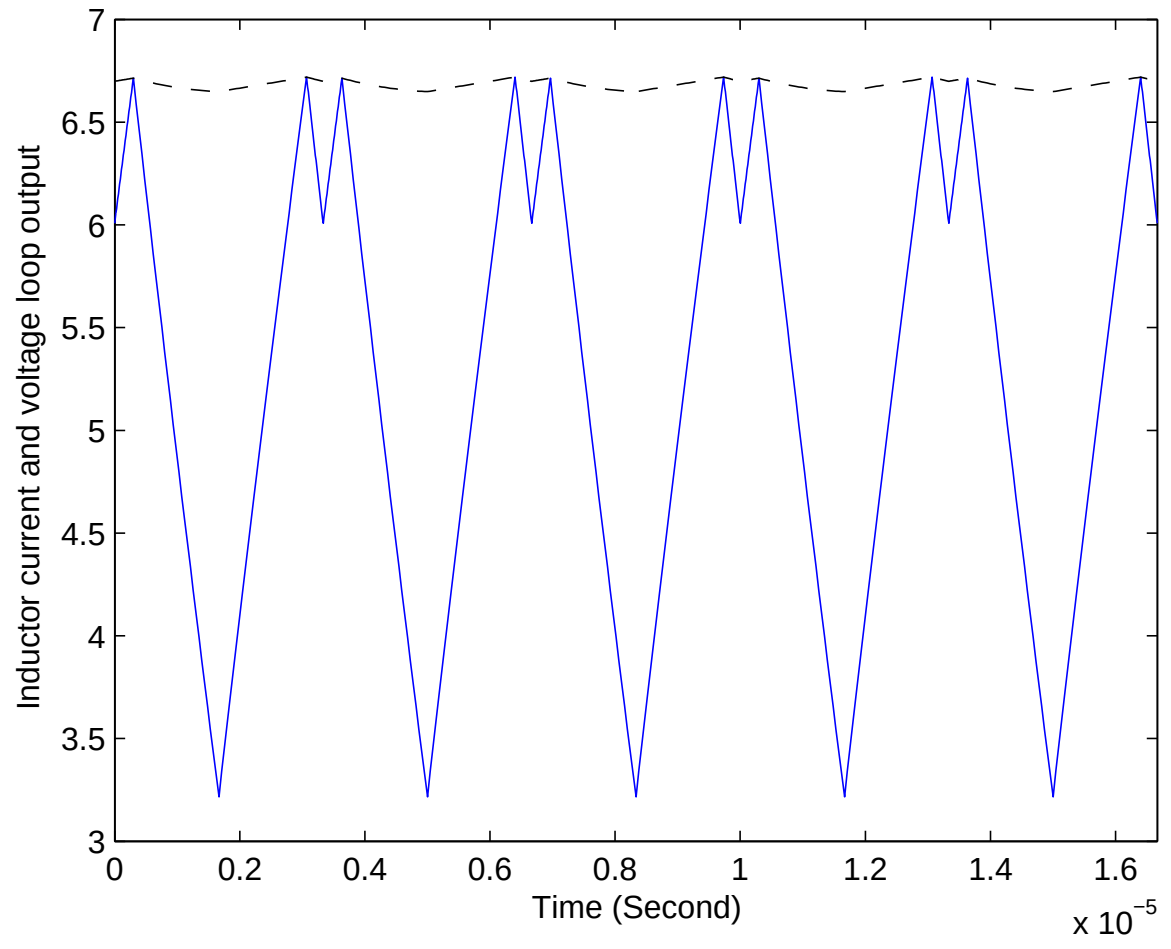

Figure 13: $2T$-periodic $i_L$ (solid line) and the voltage loop output (dashed line), $v_r = 8.4$.

## 5. Conclusion

Bifurcations associated with parasitic inductor resistance in the boost converter are analyzed, and the anslysis agrees with simulations. The critical conditions for various control schemes are summarized in Table 1. The bifurcation diagram confirms the occurrence of SNB. Both VMC and CMC are considered. Closed-form critical conditions of the bifurcations, in terms of $D$ and $v_r$, are derived. With PVMC, SNB occurs when $D = D_S$ or $v_r = v_r^*$ (see (11) and (5)). The Hopf bifurcation occurs when $D = D_H$ or $v_r = v_r(D_H)$ (see (10) and (4)). Both the bifurcations can be predicted by the average model. With a type-III compensator, SNB occurs when $D = D_S$ or $v_r \approx v_C(D_S)$.

The parasitic inductor resistance has great effects on the boost converter dynamics. If the parasitic inductor resistance is modeled, SNB occurs in VMC, or in CMC with a closed voltage loop. Either a proportional feedback or a type-III compensator is applied, SNB would occur. In real circuits, the parasitic inductor resistance always exists. If the parasitic inductor resistance is not modeled, SNB is not predicted while it actually exists, and one may be misled by the wrong dynamics and the wrong steady-state solutions.

Table 1: SNB critical conditions for various control schemes (with large feedback gains).

| **Control scheme** | **Critical condition** |
|---|---|
| **Case 1:** $r > 0$ | |
| PVMC | $D = 1 - \sqrt{r/R}$ or $v_r = v_s/2\sqrt{r/R}$ |
| VMC with type-III compensator | $D = 1 - \sqrt{r/R}$ or $v_r = v_s/2\sqrt{r/R}$ |
| CMC with voltage loop open | Generally no SNB occurs |
| CMC with voltage loop closed | $D = 1 - \sqrt{r/R}$ or $v_r = v_s/2\sqrt{r/R}$ |
| **Case 2:** $r = 0$ | |
| PVMC, VMC, or CMC | Generally no SNB occurs |

When SNB occurs, *multiple* steady-state solutions coexist. Although the fact about the possible existence of multiple solutions in the boost converter has been known, its implication and its association with SNB have not been discussed in details. Occurrence of SNB has at least two implications. First, the converter may operate with a "jump" from one solution to another. Second, among the multiple solutions, care should be taken in the compensator design to ensure that only the desired solution is stabilized [6]. When one (undesired) solution is stabilized, the other (desired) solution may be destabilized. The multiple solutions have different duty cycles.

Since SNB generally occurs in VMC, or in CMC with a closed voltage loop, one may wonder why such an instability has never been reported. In industry practice, the unstable solution with a higher duty cycle (and thus SNB) is generally prevented from existence by placing a limitation on the *maximum* duty cycle less than $D_S \approx 1 - \sqrt{\eta}$. That may explain why such an instability is rarely reported. Since $D_S$ is a function of $\eta = r/R$, one needs to know the value of $\eta$ to set the maximum duty cycle.